\documentclass[11pt,a4paper]{article}

\usepackage[T1]{fontenc}
\usepackage[utf8]{inputenc}
\usepackage{authblk,placeins}
\usepackage{xcolor}
\usepackage{listings}
\usepackage{graphicx}

\usepackage[resetlabels]{multibib}

\usepackage[margin=1in]{geometry}

\newcites{SM}{SI References}
\makeatletter
\newcommand*{\centerfloat}{%
  \parindent \z@
  \leftskip \z@ \@plus 1fil \@minus \textwidth
  \rightskip\leftskip
  \parfillskip \z@skip}
\makeatother

\title{Charting nanocluster structures via convolutional neural networks}

\author[1]{Emanuele Telari}
\author[1]{Antonio Tinti}
\author[1]{Manoj Settem}
\author[2,3]{Luca Maragliano}
\author[4]{Riccardo Ferrando}
\author[1]{Alberto Giacomello}

\affil[1]{\emph{Dipartimento di Ingegneria Meccanica e Aerospaziale, Sapienza Universit\`a di Roma, Rome, Italy}}
\affil[2]{\emph{Dipartimento Scienze della Vita e dell'Ambiente, Universit\`a Politecnica delle Marche, Ancona, Italy}}
\affil[3]{\emph{Center for Synaptic Neuroscience and Technology, Istituto Italiano di Tecnologia, Genova, Italy}}
\affil[4]{\emph{Dipartimento di Fisica, Universit\`a di Genova, Genova, Italy}}

\begin{document}
\maketitle

{\centering\texttt{antonio.tinti@uniroma1.it ferrando@fisica.unige.it alberto.giacomello@uniroma1.it}}

\begin{abstract}
A general method to obtain a representation of the structural landscape of  nanoparticles in terms of a limited number of variables is proposed. The method is applied to a large dataset of parallel tempering molecular dynamics simulations of gold clusters of 90 and 147 atoms, silver clusters of 147 atoms, and copper clusters of 147 atoms, covering a plethora of structures and temperatures.
The method leverages convolutional neural networks to learn the radial distribution functions of the nanoclusters and to distill a low-dimensional chart of the structural landscape.
This strategy is found to give rise to a physically meaningful and differentiable mapping of the atom positions to a low-dimensional manifold, in which the main structural motifs are clearly discriminated and meaningfully ordered. Furthermore, unsupervised clustering on the low-dimensional data proved effective at further splitting the motifs into structural subfamilies characterized by very fine and physically relevant differences, such as the presence of specific punctual or planar defects or of atoms with particular coordination features. Owing to these peculiarities, the chart also enabled tracking of the complex structural evolution in a reactive trajectory. In addition to visualization and analysis of complex structural landscapes, the presented approach offers a general, low-dimensional set of differentiable variables which has the potential to be used for exploration and enhanced sampling purposes.
\end{abstract}

\section{Introduction}

Finite-size aggregates -- of atoms, molecules or colloidal particles -- can present a much broader variety of structures than infinite crystals, because they are not constrained by translational invariance on an infinite lattice. 
For example, the \textit{structural landscape} of small metal particles that consist of few tens to few hundreds of atoms is much richer than that of their bulk material counterparts \cite{Johnston2002book,Alonsobook,Walesbook,baletto2005rev}. Different factors cooperate at rendering this variegated scenario: first of all, possible structures are not limited to fragments of bulk crystals, but they include non-crystalline motifs, such as icosahedra or decahedra, which contain fivefold symmetries that are forbidden in infinite crystals \cite{Nelli2023apx}. Moreover, for small sizes, also planar, cage-like, and amorphous clusters have been  observed \cite{Garzon1998prl,Bulusu2006pnas,Carles2016sr_au}, along with hybrid structures that exhibit features associated to more than one motif within the same cluster \cite{settem2022AuPTMD}.
Adding to this already complex scenario, metal nanoclusters are very likely to present defects, of which there are many different types. Volume defects for instance, such as stacking faults and twin planes, are frequently observed in experiments and simulations \cite{Dureuil2001epjd,Huang2018prm,Du2019angew,Xia2021ncomms,Elkoraychy2022nh}. Furthermore, surface reconstructions are known to occur in several clusters, \cite{apra2004AuRosette,Barnard2008jpcc,Kloppenburg2022nanoscale,Nelli2022epjap} and internal vacancies can also be stabilized in some cases \cite{Mottet1997ss,Nelli2021jcp}. Owing to the complexity of the structural landscape of nanoclusters, there is an urgent need for a robust classification method that can separate their structures into physically meaningful groups, possibly producing an informative chart of the structural landscape in terms of a small number of collective variables (CVs). In addition to providing a low-dimensional representation of the structural landscape, CVs are an essential tool of techniques to  enhance sampling in configuration space, such as umbrella sampling, \cite{torrie1974monte} metadynamics, \cite{laio2002escaping} temperature-accelerated MD \cite{maragliano2006temperature}, and many others. A common trait to most enhanced sampling approaches is the requirement that the chart be differentiable with respect to atomic coordinates, \emph{i.e.}, that the CVs are differentiable functions of the coordinates.

Machine learning (ML) is emerging as an invaluable analysis tool in the field of nanoclusters, as it allows to efficiently navigate the complexity of the structural landscape by extracting meaningful patterns from large collections of data. ML has already found application in microscopy image recognition \cite{palmer2022}, dimensionality reduction and exploration of potential energy surfaces \cite{PES}, structural recognition \cite{roncaglia2023,PES,anker2022extracting}, characterization of the local atomic environment \cite{filion2019,baletto2021}, and machine learnt force fields for metals \cite{zeni2018building}.

One of the main challenges in the study of nanoclusters concerns the identification of descriptors that can discriminate the various structural classes. The availability of such a tool is crucial for navigating the landscape of structures generated during simulations. In this context the histogram of the interatomic distances, \emph{i.e.} the radial distribution function (RDF), has been used to study the solid-solid transitions in metallic/bimetallic clusters via metadynamics,\cite{pavan2015NPsMeetMetaD} owing to its capability to encode structural information. Another widely used approach, is Common Neighbor Analysis (CNA) \cite{cna1994}, a tool which relies on analyzing  local atomic coordination signatures for individual atoms \cite{schebarchov2018AuHSA,roncaglia2023}. Often, arbitrary rules \cite{schebarchov2018AuHSA,settem2022AuPTMD} are then applied to CNA signatures of the atoms as a means to assign the whole nanocluster to a structural family. Albeit being widely used and informative, CNA still presents certain drawbacks. First, CNA classifications are based on the arrangement of first neighbors around any given atom, and therefore they do not directly encode information on the overall shape of the nanoparticles. In addition, even though CNA can be used for charting the structural landscape and for unsupervised clustering to obtain very refined groupings of structures (\emph{e.g.}, along the lines developed by Roncaglia \& Ferrando \cite{roncaglia2023}), the resulting chart is non-differentiable.

In this work, we propose to use a descriptor capable of capturing in full generality the most important structural features of metal nanoclusters --the RDF-- and feed it to an artifical neural network (ANN) that is trained to perform an unsupervised dimensionality reduction, yielding a low-dimensional, informative representation, where data are distributed according to their structural similarities. We start off by showing that RDFs are excellent descriptors of nanocluster structures, given their capability to describe both the local  \cite{delgado2021universal} and global order together with the overall shape of diverse systems, and then we proceed to  discuss the results obtained by using convolutional ANNs to reduce the dimensionality of the original descriptors. 

The combination of RDF and ANNs allowed us to learn a differentiable map from the atomic positions to a low-dimensional (3D) chart of the structural features of nanoclusters of various sizes and metals. The employed datasets contain hundreds of thousands of unique structures obtained by parallel-tempering molecular dynamics (PTMD) simulations \cite{settem2022AuPTMD,settemAgCu}. It was possible to classify in an unsupervised manner this wealth of structures, reproducing the well-known CNA classes and, additionally, being able to distinguish subtle features present in metal nanoclusters, including location of the twinning planes stacking faults, surface defects, central vacancies in icosahedra, and intermediate/distorted structures. The chart also allowed us to track and describe in detail dynamical structural transformations.
Additional advantages of the present chart are the transferability and robustness, which was demonstrated using independent datasets of metal clusters of varying size and chemical nature, together with its differentiability (and hence suitability for CV-based exploration and biasing in molecular dynamics).

\section{Results and discussion}
Our goal is to gain insights into the structural complexity of metal nanoclusters by means of a differentiable map of the of configuration space onto a low-dimensional, yet sufficiently informative manifold (the chart).

The method consists in generating, for every cluster configuration in the dataset, a set of high dimensional descriptors, the RDFs, which are known to describe both the local structural order and global shape, and distill this information representing it in a low-dimensional, highly compressed form.
The specific ANN architecture we chose to perform the unsupervised dimensionality reduction is that of an autoencoder (AE) \cite{hinton2006autoencoder} endowed with convolutional layers that renders it highly specialized at learning from numerical sequences \cite{kiranyaz20211d}.  A dimensionality reduction step follows the convolutions, yielding a physically informed three-dimensional (3D) chart of the structural landscape of our dataset, which allows to navigate and easily understand it. Finally, we apply a clustering technique to the 3D chart to gauge its quality and to identify different structural families.

\begin{figure}
        \centering
        \includegraphics[width= 1.0\textwidth]{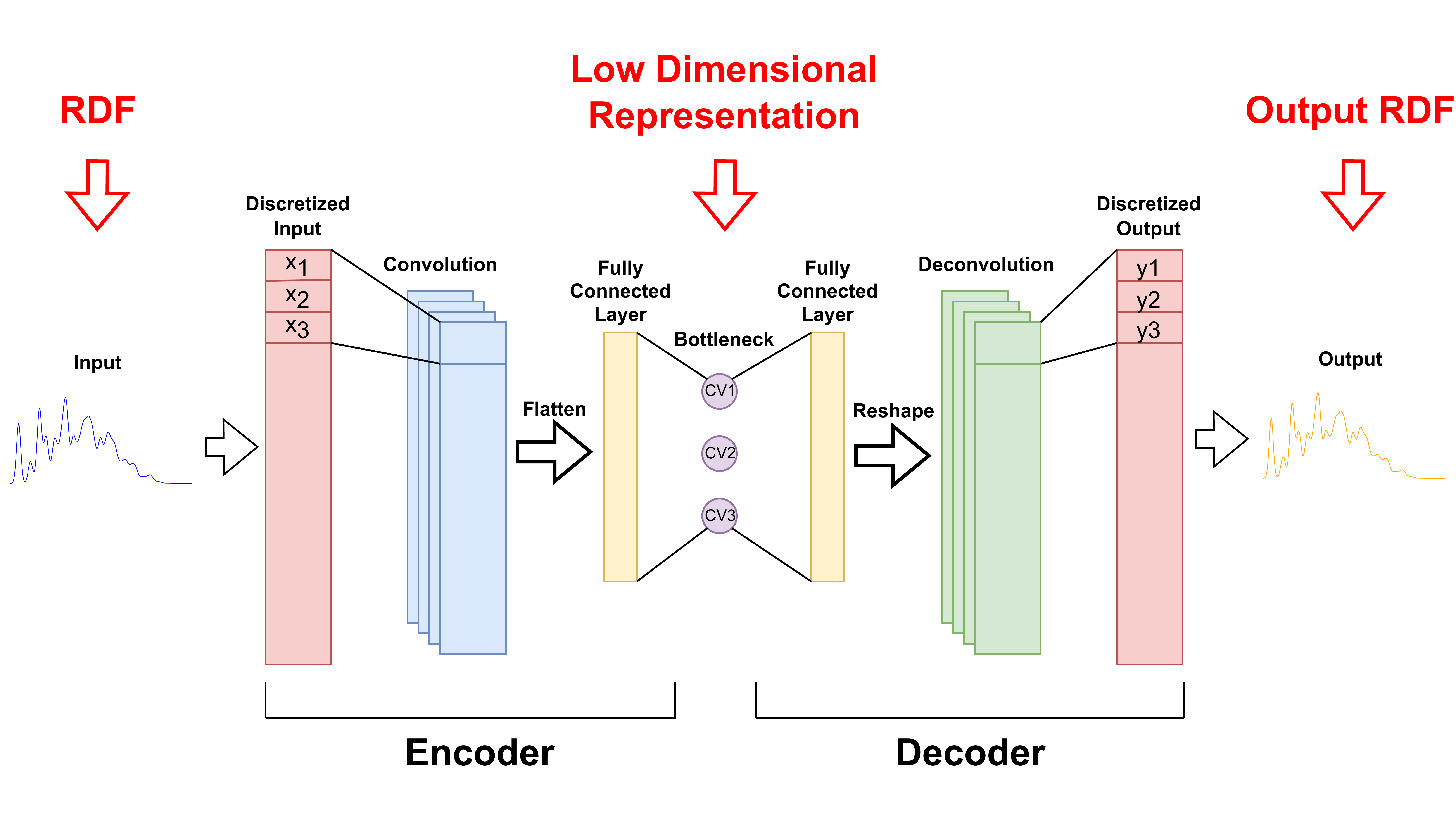}
        \caption{Simple sketch of the autoencoder architecture, showing how encoder and decoder meet at a low-dimensional (3D) bottleneck.}
        \label{fig:CNNsketch}
\end{figure}

AEs constitute a particular class of ANNs that is highly specialized in the unsupervised dimensionality reduction of data \cite{hinton2006autoencoder}. AEs are designed to reproduce the input while forcing the data through a bottleneck with severely reduced dimensionality (Fig.~\ref{fig:CNNsketch}). In this way, the network needs to learn, in the first section of the network (encoder), an efficient representation of the data in such a way that the information can then be reconstructed by the second half of the newtork (decoder) with sufficient accuracy. The quality of the reconstruction, is measured by a loss function that is also used in the training the network.

Convolutional layers, which are specialized at learning from ordered sequences, are adopted in the AE hereby presented, because discretized RDFs are by all means sequences.
They work applying different kernels that slide along the data allowing the recognition of local features and patterns, which makes them well versed for the analysis of inputs like signals (using 1d convolutional kernels) or images (2d kernels). Moreover, the connections between the nodes and the related parameters are considerably reduced as compared to the fully connected layers used in standard ANN, which decreases the computational cost while allowing for better performances.

In order to test the method, we took advantage of the large dataset of nanocluster structures produced by the group \cite{settem2022AuPTMD,settemAgCu} via parallel tempering molecular dynamics (PTMD) for gold, silver, and copper nanoclusters of different sizes. 
In the next section we discuss in detail the results obtained for the most challenging case --a gold cluster of 90 atoms, Au$_{90}$-- while results relative to other metals and sizes will be shown in later sections.

\subsection{Structural landscape of Au$_{90}$}
Gold nanoclusters represent an ideal test case, owing to the broad variety of structures \cite{Garzon1998prl,Bulusu2006pnas,Carles2016sr_au,settem2022AuPTMD,apra2004AuRosette} they present, which include face-centered-cubic (fcc) lattice, twins, icosahedra (Ih), and decahedra (Dh).  In the following, nanoclusters will be broadly classified into such standard structural families by CNA (in addition to the mix and amorphous classes), as used by Settem et al. \cite{settem2022AuPTMD},  with the aim of having an independent benchmark for our unsupervised study.
Here we focus on a small gold nanocluster, Au$_{90}$, which is characterized by an extremely challenging structural landscape, owing to the large fraction of surface atoms.
In particular, we chart a set of Au$_{90}$ configurations  extracted from PTMD simulations \cite{settem2022AuPTMD} exploring a total of 35 temperatures ranging from 250~K to 550~K. 
Starting from an initial set of 921,600 atom configurations, we performed a local minimization and filtered out duplicates, reducing the dataset to 49,016 independent configurations.

\begin{figure}
        \centering
        \includegraphics[width= 1.0\textwidth]{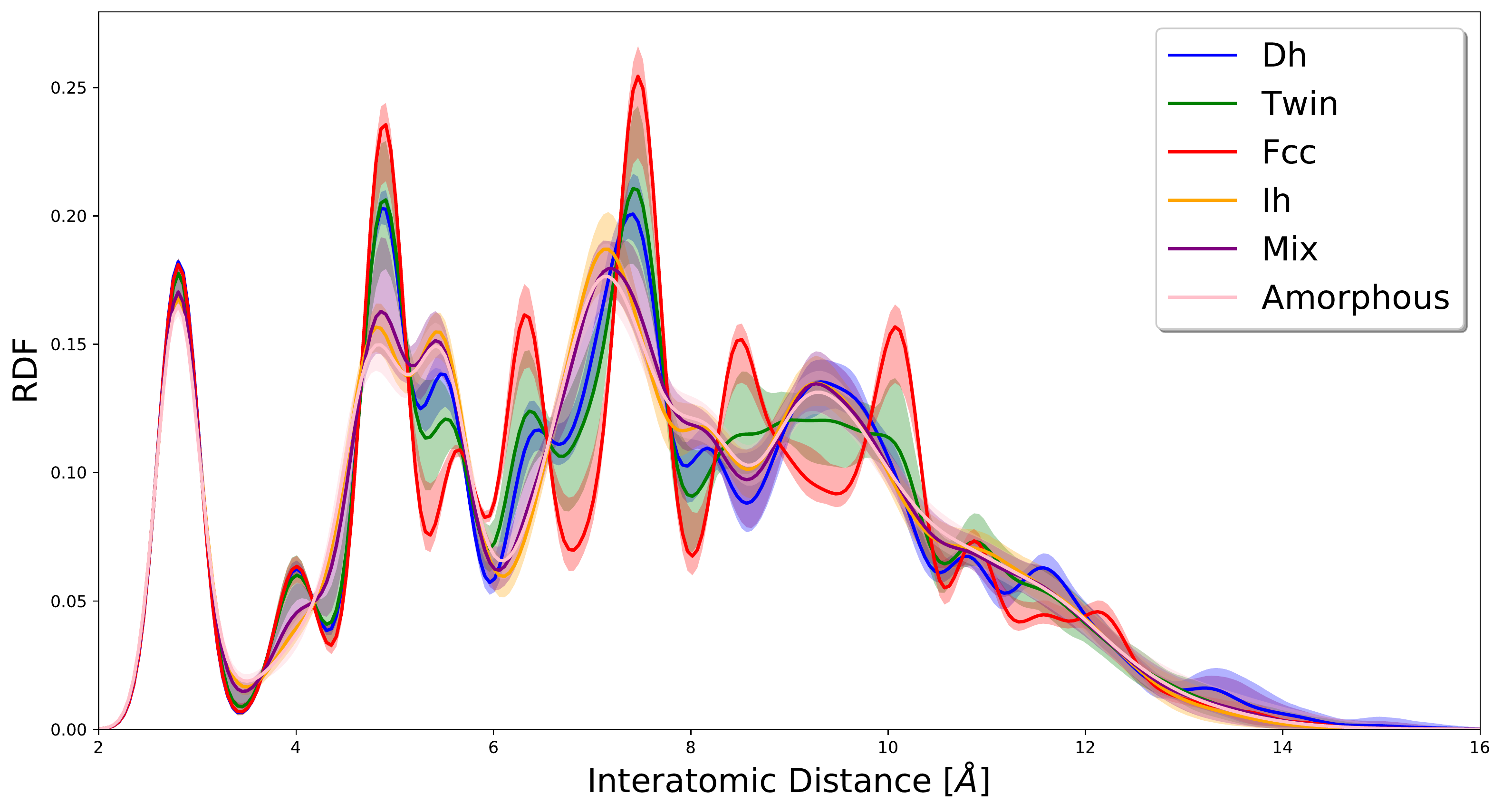}
        \caption{Radial distribution functions families for  Au$_{90}$. Colors reflect cluster structure classification provided by CNA. Blue is used for Dh, green for Twin, red for Fcc, orange for Ih, purple for Mix, and pink for Amorphous. Shaded areas represent intervals containing 90 \% of the data for each CNA label, with the lower boundary representing the 0.05 quantile of the RDF population and the upper boundary the 0.95 quantile.}
        \label{fig:rdfsau}
\end{figure}

As  previously mentioned, RDFs were chosen because they are general descriptors of short and long range order \cite{hansenMcdonald,christiansen2020} that are equivariant with respect to rototranslation and permutation of the atom coordinates. The aptness of RDFs as structural descriptors is well demonstrated by Fig.~\ref{fig:rdfsau}, in which the RDFs of all CNA classes (fcc, twin, Dh, Ih, mix, and amorphous) are well separated. We will show in the following that this descriptive power also applies to other metals and nanocluster sizes, which actually have a less rich structural landscape. However, a major drawback of using a probability distribution as a descriptor --even in its discretized version-- is its high dimensionality. Our approach to provide an efficient charting of the  structural landscape of metal nanoclusters, \emph{i.e.}, a low-dimensional representation, relies therefore on a dimensionality reduction step.

A large number of RDFs, corresponding to individual PTMD-derived structures, are used to train an autoencoder (AE), which automatically learns to compress the high-dimensional RDF information to a 3D latent representation (Fig.~\ref{fig:CNNsketch}).
Our AE is composed by an input and an output layer, a central block, comprising the bottleneck layer, formed by three fully-connected layers, while the cores of the encoder and the decoder are formed by convolutional layers (Fig.~\ref{fig:CNNsketch}). The training was run feeding  the AE with the  RDFs dataset (49,016 independent data), split in training and validation sets; the mean squared error (MSE) between the output and the input RDF is used as the loss function.
We chose to adopt a latent space dimensionality of 3. This choice allowed for better performances in terms of the loss function as compared to higher compressions, while still allowing for a convenienent visual representation. We refer to the Supporting Information for a comparison of the results obtained varying the dimensionality of the latent space.

\begin{figure}
        \centering
        \includegraphics[width= 1.0\textwidth]{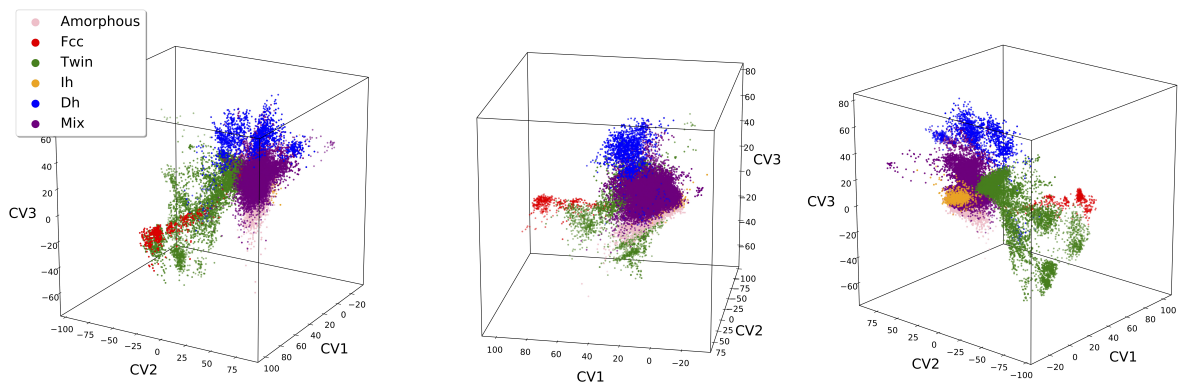}
        \caption{Visualization of the 3D chart generated via convolutional AE for Au$_{90}$ dataset, from different perspectives. Individual points refer to a given Au$_{90}$ configuration in the dataset mapped according to their latent space representation. The three latent coordinates are referred to as \emph{CVs}.
        Points are colored following their (independent) CNA label classification; the color code is the same used in Fig.~\ref{fig:rdfsau}. 
        }
        \label{fig:chartau90}
\end{figure} 

The 3D chart obtained by the AE is shown in Fig.~\ref{fig:chartau90} with datapoints colored by their CNA label. This representation clearly indicates how each structural family is grouped in separate regions of the chart and how their spatial ordering and distance reflects affinities among these families: similar structures are placed close together (\emph{e.g.}, fcc and twin), while structures that share common features occupy intermediate regions (\emph{e.g.}, the twin region is interposed between fcc and Dh). 
Overall, the obtained chart allows for a physically meaningful representation of the structures. The scatter in the data suggests that the resolution of the analysis of the chart allowed by the CNA summary labels is not fully conclusive and how further analysis can allow for a better understanding of the physical information encoded in the structures distribution inside the latent space and, consequently, a finer discrimination of different families of structures.

\begin{figure}
        \centering
        \includegraphics[width= 1\textwidth]{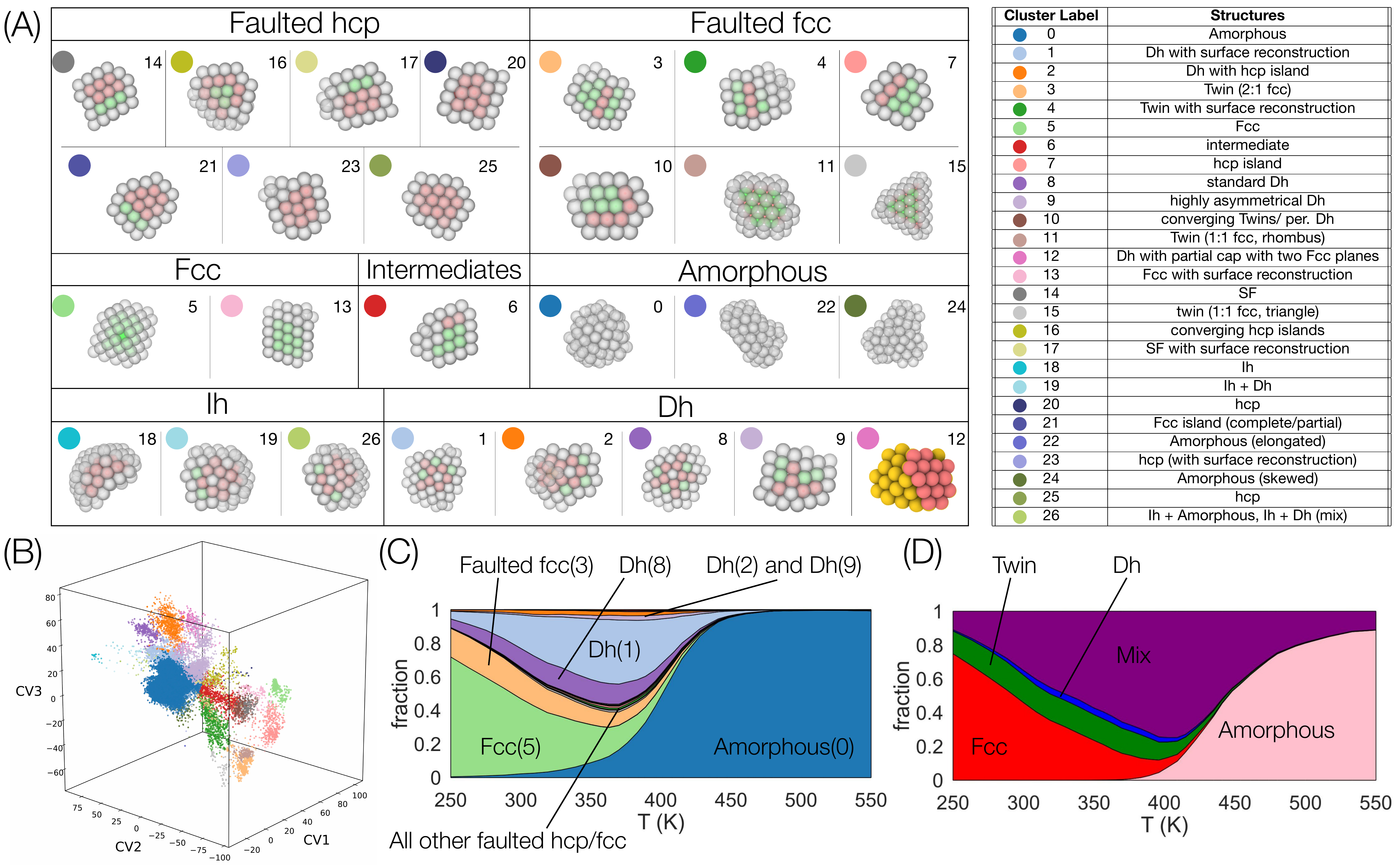}
        \caption{A) Representative samples for each of the 27 structural families identified via application of the mean shift clustering algorithm on the latent space representation of the Au$_{90}$ dataset. These 27 classes were subsequently grouped in 7 bigger families by similarity. Atom colors refer to their coordination: green represents atoms with fcc coordination, red stands for hcp coordination, white for neither of the previous ones.  Atomistic representations with transparency report 3D views, whereas those in solid colors represent cross sections. Every structure is given a numeric index associated with the label of the belonging cluster and a particular the color. The table on the right reports both the numeric and color labels of the clusters along with a description of the various structures.
        B) Single view for a 3D plot, analogous to the one on the extreme right of Fig. \ref{fig:chartau90} except for the coloring, which is now representative of the labels assigned by the mean shift through the same color coding reported in panel A. 
        C) Mean shift families fractions as a function of the temperature in the whole PTMD dataset. The color code is the same of panels A and B. More likely structures are represented with the same name of the macro-family, numeric index and color of panel A.
        D) Plot analogous to panel C with the only difference that the PTMD data as been classified using the CNA label classification as in the work of Settem et al. \cite{settem2022AuPTMD}. Color code and labeling are the same used in Fig. \ref{fig:rdfsau}.}
        \label{fig:ms27au90}
\end{figure}

In order to increase the structural resolution and to have deeper insights into the physical information encoded in the latent space, we applied a clustering technique to identify meaningful and coherent regions in the chart. In particular, we chose a non-parametric technique known as mean shift \cite{MeanShift}. Applying this method to the 3D chart of Fig. \ref{fig:ms27au90}, was justified not only by the non-parametric nature of the clustering technique but also by its aptness at dealing with clusters of different sizes and shapes. The only input variable required by mean shift is the bandwidth, which dictates the resolution of the analysis, with the smaller bandwidths leading to more detailed parceling of the data. We chose a bandwidth that yields a robust clustering of the chart with sufficient detail as discussed in the Supporting Information. Our analysis resulted in a robust discrimination of 27 major regions for the Au$_{90}$ chart, corresponding to 27 different major structural families, as reported in Fig.~\ref{fig:ms27au90}. From the figure it is immediately apparent how the mean shift classification is able to distinguish and split clusters that belong to spatially separated regions of the chart, properly reflecting the ordering of the data.

Representative structures of each mean shift family are shown in Fig.~\ref{fig:ms27au90}A, while Fig. \ref{fig:ms27au90}B shows the 3D chart with the points colored according to the same families. They are broadly categorised into  Ih, Dh, fcc, faulted fcc, faulted hcp, intermediates, and amorphous.  Faulted fcc nanoclusters are those with a predominant fcc part but which contain twin planes and/or stacking faults.  Faulted hcp clusters are those with a predominant hcp part but which contain twin planes and/or stacking faults.
Typically, structures observed in experiments and simulations are classified into basic structural families \cite{wang2012AuImaging,wells2015AuImagingACFraction,foster2018AuImagingACFraction,schebarchov2018AuHSA,settem2022AuPTMD} which rarely capture the fine geometrical details within a given family. In contrast, our approach leads to a physically meaningful classification along with capturing the fine structural details, by splitting the broader families into several subfamilies. A closer look at the various fcc and hcp faulted nanoclusters illustrates this point.  There are three subfamilies (cluster-3, cluster-11, cluster-15) which contain only one hcp plane. Cluster-3, referred to as 2:1 fcc, consists of two and one fcc plane(s) on either side of the hcp plane. Similarly, clusters-11, 15 are 1:1 fcc with differing shapes. When the hcp plane is adjacent to surface layer, we have hcp islands (clusters-7). Cluster-10 has two converging hcp islands. In cluster-4, local surface reconstruction occurs along with a single hcp plane. Moving on to faulted hcp structures, three hcp planes converge in cluster-16. With the increase in the number of parallel hcp planes, we have either stacking faults (cluster-14) or fcc island (cluster-21) which contains one fcc plane (opposite of hcp island). In the extreme case, we have full hcp particles (clusters-20, 25). Clusters-17 and 23 both undergo local surface reconstruction similar to cluster-4.

In fcc families, we have the  conventional fcc structures (cluster-5) and fcc structures with local surface reconstruction (cluster-13). In the case of decahedra, there are five sub-families. Clusters-8, 9, and 12 are all conventional decahedra. In cluster-9, the decahedral axis is at the periphery as opposed to clusters-8 and 12. Additionally, cluster-12 has a partial cap on top (atoms belonging to the cap are shown in red color). Decahedra in cluster-2 have an hcp island on the surface. Finally, decahedra also exhibit reconstruction at the reentrant grooves resulting in icosahedron-like features (cluster-1). There are three icosahedral clusters: Cluster-18 consists of incomplete non-compact icosahedra; cluster-19 is combination of Ih and Ih+Dh (has features of both Ih and Dh) while cluster-26 is a combination of Ih+dh and Ih+amor (has features of both Ih and amorphous). Similarly, there are three types of amorphous structures (clusters-0, 22, and 24). Finally, we have intermediate structures in cluster-6.

The structural distributions of Au$_{90}$, \emph{i.e.}, the fraction of various families as a function of temperature, of the PTMD data according to mean shift and CNA labels are shown in Figs. \ref{fig:ms27au90}C and D, respectively. In both cases, we find the conventional structure families. However, mean shift further refines the CNA-based classification \cite{settem2022AuPTMD}. For instance, with mean shift, we have a clear separation of the various types of Dh that were previously grouped together in a broad group of mixed structures. In the case of faulted structures, there is a prominent faulted fcc cluster (Faulted fcc-3) while all other faulted structures (band between Faulted fcc-3 and Dh-8 in Fig.~\ref{fig:ms27au90}C) have very low fractions. It is noteworthy that mean shift can classify even structures that have very low probability of occurrence.

In short, the Au$_{90}$ analysis showcased the descriptive power of RDFs and the capability of the unsupervised dimensionality reduction performed by AE to properly compress information. Through the AE we were able to generate a highly physical representation of the data, which, rather than simply splitting different structures, is able to coherently distribute them in a 3D chart according to their physical similarities.
As a consequence, the subsequent independent classification via mean shift easily identified a wealth of distinct structures and underscored the capability of the approach to distinguish both local and global structural motifs: location of twinning planes, surface defects, distorted cluster shapes, etc.

\subsection{Generality of the approach}
\begin{figure}[h!]
        \centering
        \includegraphics[width= 1\textwidth]{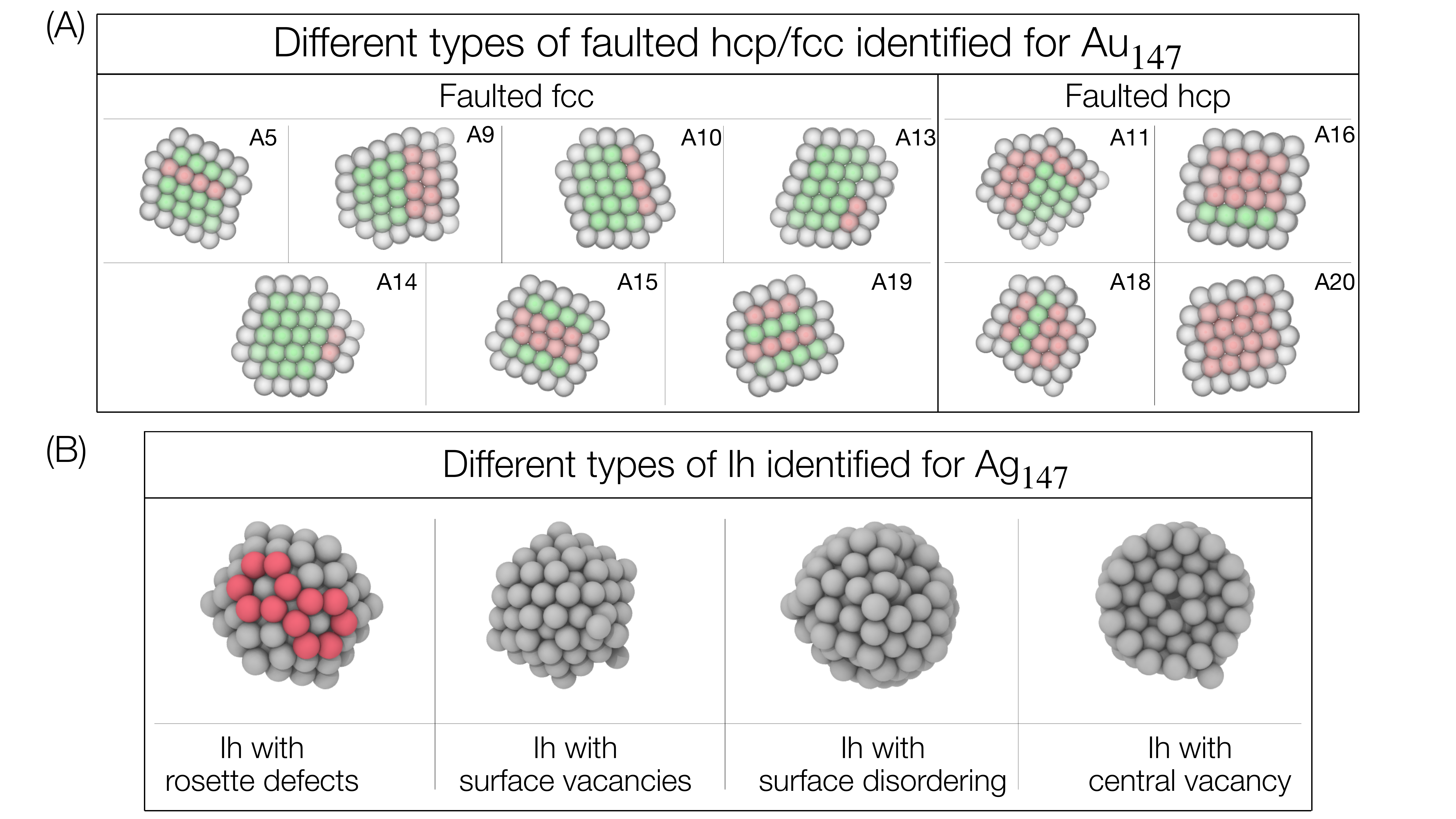}
        \caption{A) Cross-sections of the different types of twin families obtained by using mean shift clustering on the latent space representation of Au$_{147}$. The families were splitted in two groups,in the same fashion of our treatment for the Au$_{90}$ twin structures. 
        Colors of the atoms refer to their individual coordination, similarly to Fig. \ref{fig:ms27au90}.
        Every structure is labeled with the same alphanumeric index of Fig. \ref{fig:unbIhDh}A, where the 3D chart of Au$_{147}$ is depicted
        B) The four different families of icosahedral structures for Ag$_{147}$ are sketched. As customary the families were extracted via mean shift clustering in the 3D space resulting from encoding of the Ag$_{147}$ dataset. The six-atom rosette defects are highlighted in red. The first three figures on the left are three dimensional representations, the last figure is a cross section. Complete description of the clustering of all the Ag$_{147}$ structures can be found in the Supporting Information.}
        \label{fig:au147ag147}
\end{figure}
In this section we show that the approach adopted for Au$_{90}$ is of general applicability. At the root of such generality is the wealth of structural information carried by RDFs, which are expected to be valuable for a broad class of systems which includes nanoclusters of other metals and sizes, as showcased below, but is not limited to them \cite{Walesbook,filion2019}. 

Here we focus on larger cluster sizes that, as a general trend, show a lower variety of structures as compared to smaller ones. In particular, we study clusters of $147$ atoms with elemental  gold (Au$_{147}$), copper (Cu$_{147}$), and silver (Ag$_{147}$). These two latter cases exhibit rather different properties as compared to the gold clusters; in particular, they exhibit a lower differentiation in the structural landscape that is mainly dominated by Ih structures. 
We discuss only selected structural families identified by the method for the three cases, that best showcase the discerning capabilities of the method: faulted structures characteristic of Au$_{147}$ and on the different types of Ih present in Ag$_{147}$. Results for Cu$_{147}$ are similar to Ag$_{147}$ and are reported in the Supporting Information. These two examples put our approach to a test, because these two families are characterized by distinct structural features: faulted structures mainly differ for small changes in the overall shape of the particles and for their atomic coordination while Ih have more similar shapes and lower degrees of crystallinity. 

Figure~\ref{fig:au147ag147}A shows that, in the case of Au$_{147}$, our approach is capable of distinguishing fine features in the large family of faulted structures, which are broadly grouped into faulted fcc and faulted hcp, in analogy to Au$_{90}$. In the standard faulted fcc (A5, corresponding to a standard double twin), there is a single hcp plane with at least one fcc plane on either side. When the hcp plane is adjacent to the surface layer, we have hcp islands (A10) or sometimes partial hcp islands (A13, A14). In addition, an hcp plane and an hcp island can occur within the same structure (A19). When there are more than one hcp plane, stacking defects are observed. In the extreme case, it can be completely hcp (A20) or fcc island (A16). When there are two hcp planes, depending on the location of the hcp planes, we have either the central stacking fault (A15) or peripheral stacking fault (A9). In the standard faulted hcp (A18), there is a single fcc plane with at least one hcp plane on either side. Finally, we have the faulted hcp cluster with converging hcp planes (A11).

Owing to the particular characteristics of silver, the structural landscape of Ag$_{147}$ is largely dominated by icosahedra, which the clustering method is able to split into four subfamilies (Fig. \ref{fig:au147ag147}B). Conventional Ih consisting of surface vacancies are the dominant among them. Icosahedra also undergo reconstruction and disordering through ``rosette'' defects on the surface. When the disordering increases further, we observe Ih with surface disordering. Finally, one can recognize Ih with a central vacancy where the central atom is missing as shown in the cross section in the rightmost panel of Fig.~\ref{fig:au147ag147}B. Distinguishing with ease the latter structural subfamily is a feature of our approach; indeed CNA can hardly recognize icosahedra with a central vacancy because it relies on the (missing) Ih-coordinated atom to identify the Ih class.

In summary, for all the considered cases, the method proved to be transferable and robust, being capable of  characterizing the wealth of structures of Au$_{147}$ and giving insights into the fine features distinguishing Ih subclasses for Cu$_{147}$ and Ag$_{147}$.

\subsection{Dynamical structural transitions}

\begin{figure}[h!]
        \centering
        \centerfloat{\includegraphics[width= 1.\textwidth]{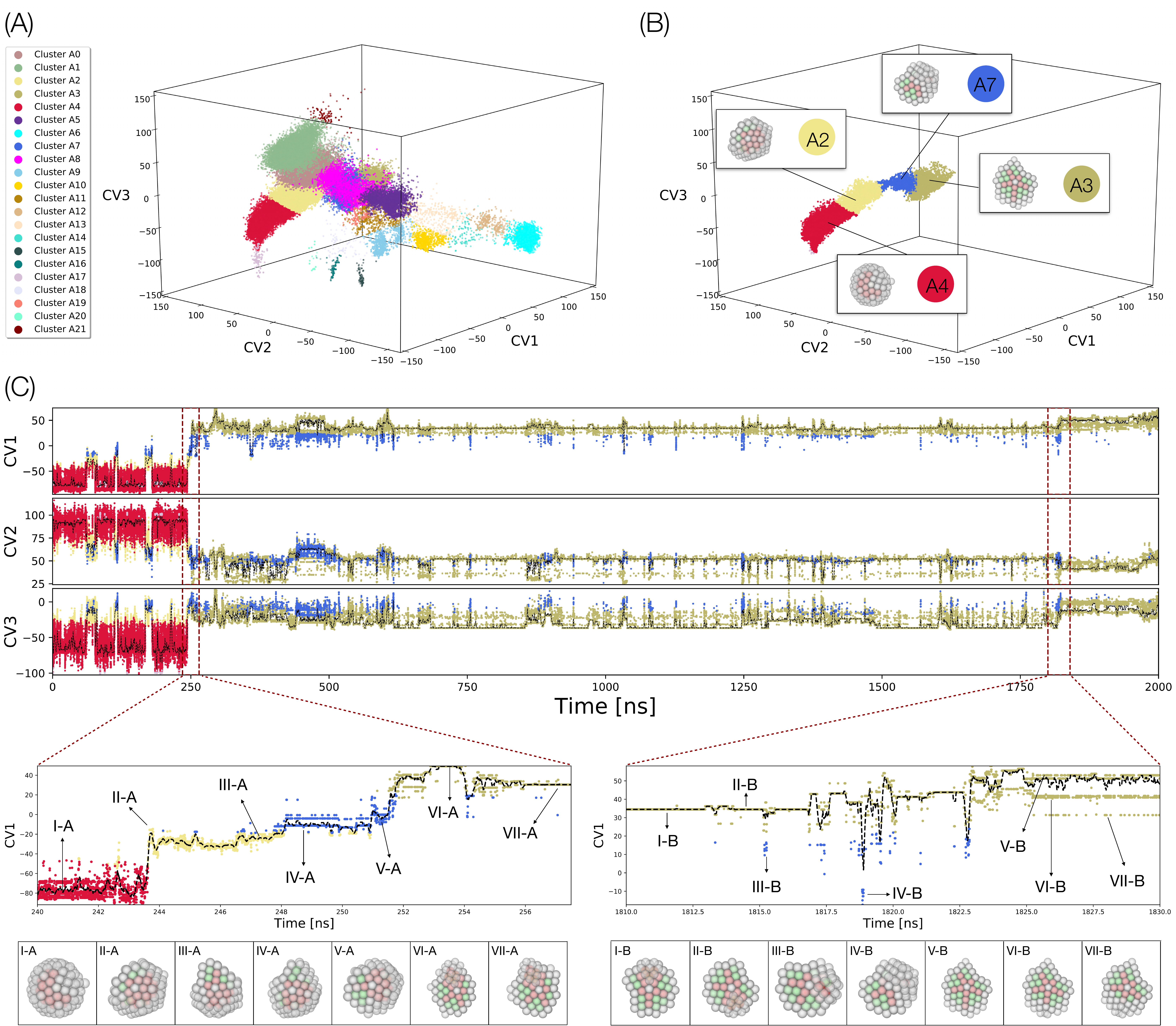} }
        \caption{
        A) Structural chart of Au$_{147}$ containing 87,050 structures.
        Points are colored according to the structural families identified by mean shift clustering, see also Fig.~S4, now labeled using alphanumeric indexes to distinguish them by the families of Fig.~\ref{fig:ms27au90}.
        B) Plot of an unbiased MD simulation of Au$_{147}$ undergoing a structural transition from Ih to Dh in the same chart as A.  
        The point are colored using their mean shift classification obtained on the training dataset represented in panel A. In the plot are depicted representative structures of the different regions.
        C) Scatter plots of the time evolution of the three CVs along the trajectory of panel B. Dark red dashed lines highlight two intervals in which the main transformations from Ih to Dh occurs. The colors of the points correspond to their mean shift label as in panel A and B. Black dashed lines represent a running average of the scatter plots. Bottom panels report magnifications of the two main transitions with snapshots of the main structures observed. 
        }
        \label{fig:unbIhDh}
\end{figure}

The previous sections demonstrated how the method at hand is capable of generating reliable, low-dimensional structural charts from large datasets of nanoclusters configurations for different metals and sizes. 
In all considered cases, the charts, informed by RDFs, excelled at distributing the different families of structures in a physically meaningful fashion, keeping similar structures closer while positioning different ones far apart.
The method was able to distinguish both structures presenting major shape differences (as faulted fcc and hcp in Au nanoclusters) and structures with lower degrees of crystallinity and closer overall shape (Ih subfamilies).
In other words, the three CVs defining the chart can discriminate between different metastable states of the systems studied while mantaining an insightful ordering among them. These features suggest that the approach can be used for describing  structural transitions occurring along reactive trajectories, \emph{e.g.}, obtained by MD simulations.
To test this idea, we use the chart to study a continuous dynamical trajectory (Fig.~\ref{fig:unbIhDh}).

We consider a 2 $\mu$s unbiased MD run of Au$_{147}$ at $396$~K. At this temperature, the most probable structure for Au$_{147}$ is Dh \cite{settem2022AuPTMD}. By choosing as  initial configuration an Ih structure, which is very unlikely in such thermodynamic conditions, it is possible to observe a spontaneous Ih $\to$ Dh transition in an unbiased trajectory. In particular, we map 2 millions of individual MD snapshots on the chart through the AE in Fig.~\ref{fig:CNNsketch}, which was previously trained on independent structures generated by PTMD. To be compatible with this representation, each snapshot undergoes a short local  minimization.

Figures~\ref{fig:unbIhDh}A, B compare the structural chart of the entire PTMD dataset with the partial representation of the same chart as obtained from the unbiased MD trajectory.  
The trajectory progressively populates a connected, tube-shaped region of the chart, which joins smoothly Ih to Dh domains, passing through intermediate, defected structures which belong to well defined families. More in detail, the following structural pathway is observed: Ih (cluster-4) $\rightarrow$ distorted-Ih (cluster-2) $\rightarrow$ distorted-Dh (cluster-7) $\rightarrow$ Dh (cluster-3) which is confirmed by analyzing the structures along the trajectory (Fig.~\ref{fig:unbIhDh}C). 
Beginning from Ih there is an initial transition to distorted-Ih where the disorder increases and we start observing fcc-coordinated atoms in the nanocluster. The distorted-Ih then changes to distorted-Dh where the amount of fcc coordinated atoms increases further. Apart from the difference in the amount of fcc, distorted-Ih is geometrically similar to Ih while distorted-Dh is closer to Dh. Finally, the distorted-Dh transitions to Dh which completes a gradual change from Ih to Dh with physically meaningful changes along the tube-shaped region.

In the absence of the chart, it would in principle be possible to perform a visual analysis of the Ih $\rightarrow$ Dh trajectory of roughly two millions of structures. However, it would be extremely cumbersome to identify the main thermally activated transformation, and to track the fine structural changes and fluctuations along the trajectory which are crucial for understanding the transition mechanisms. This difficulty is easily overcome by tracking changes in the chart coordinates as reported in  Fig.~\ref{fig:unbIhDh}C, which shows the time evolution of the CVs as a function of time along the trajectory. Changes in CVs are found to correlate very well with structural changes. Three broad phases can then be distinguished during the evolution of the trajectory. In the initial phase (up to $\sim$ 250 ns), the nanocluster is predominantly Ih (cluster-4) with intermittent fluctuations to distorted-Ih (cluster-2), distorted-Dh (cluster-7). The actual Ih$\to$Dh transition occurs around $\sim$ 245 ns, followed by a long intermediate phase (spanning $\sim$ 245 ns to $\sim$ 1820 ns), in which fluctuations between  Dh (cluster-3, dominant) and distorted-Dh (cluster-7, minor) are observed. A final transition step at $\sim$ 1820 ns leads to the final phase  consisting of Dh with very few fluctuations to distorted-Dh. Here, we stress that this information can be obtained simply by following the CVs even before analyzing the structures.

We will now focus on the transition regions and look closely at the structural changes. For this purpose, we consider CV1. In the tube-like region, a continuous increase in CV1 is synonymous with a continuous change from Ih to Dh. A zoomed plot of the first transition (between 240 ns and 260 ns) is shown in the lower left panel of Fig. \ref{fig:unbIhDh}C, see Fig.~S7 for CV2 and CV3. The initial Ih structures (I-A) transition to distorted-Ih structures (II-A, III-A) where we begin to see the fcc-coordinated atoms along with Dh-like features. With further increase in CV1, there is a gradual change to distorted-Dh structures (IV-A, V-A). Finally, these structures transition to Dh structures which have an hcp island (VI-A, VII-A). Decahedra with hcp island dominate the middle phase and hcp island-free Dh are obtained after a final transition around $\sim$ 1822 ns (shown in the lower right section of Fig.~\ref{fig:unbIhDh}C). This second transition is marked by a slight increase in the mean CV1 value (black dashed line): initially, we have Dh with hcp island (I-B, II-B) which transition to a better Dh (without hcp island) around $\sim$ 1823 ns (V-B). It appears that this transition is aided by fluctuations to distorted-Dh intermediates (III-B, IV-B). After the transition to a better Dh (beyond $\sim$ 1825 ns), there are three distinct horizontal branches. The dominant one, which has the highest CV1 value, corresponds to the perfect defect-free Dh (V-B). However, this structure often undergoes two types of local reconstructions near the reentrant groove (VI-B, VII-B), which  coincide with two distinct values of CV1.

The preceding discussion underscores that the three deep CVs are capable of describing in a detailed and physical fashion what happens during a dynamical transition. The chart enables on-the-fly tracking of the system along its structural changes and describes transitions between different metastable states.
This is a further evidence of the physical insightfulness of the latent space generated starting from the RDFs, underscoring the reliability of the structural information contained in the charts and further showcasing the power of the approach. 
In particular, the method shows promise for  characterizing and analyzing long trajectories generated via molecular simulations enabling a fast and informed way to study and follow the time evolution of this type of systems. Importantly, the differentiability of the coordinates of the latent space with respect to the atomic positions opens the way to  address the challenge of biasing MD simulations of structural changes \cite{pavan2015NPsMeetMetaD,tribello2017}. The specific merit of this approach is to provide a natural route to devise a general, informative, and low-dimensional collective variable space capable of describing dozens of structural motifs. We plan to investigate  structural transformation driven by deep learnt collective variables in a separate communication.

\FloatBarrier

\section{Conclusions}\label{sec13}

This work presents an original machine learning method capable to chart the structural landscape of nanoparticles according to their radial distribution function. The approach comprises two subsequent information extraction steps. The first consists in translating the atomic coordinates into RDFs, which encode information about the structure in translationally, rotationally, and permutationally invariant way. The high dimensional information contained in the RDF is then reduced to a low-dimensional (3D) and yet informative representation  (``chart'') by exploiting convolutional autoencoders. These deep-learnt collective variables are surprisingly good at describing structural features in a physically meaningful way, discriminating the different states of the system.

The 3D charts of different metal nanoclusters were then analysed using a non-parametric clustering technique, which allowed us to classify the datapoints into structural families. 
The method succeeded at disentangling the complex structural motifs of nanoclusters having different shapes and metals (Au$_{90}$, Au$_{147}$, Ag$_{147}$, and Cu$_{147}$), distinguishing also fine difference between faulted and mixed structures as well as small defects (icosahedra with central vacancy, surface defects, etc.). Related structural motifs, \emph{e.g.}, fcc and faulted fcc/hcp were found to occupy close regions of the chart, allowing us to garner insights also into dynamical structural transformations.

Finally, the method further proved useful in  the analysis of a long unbiased MD run of Au$_{147}$ undergoing a structural transition. The collective variables allowed us to accurately  track and describe structural changes along the dynamics. This pushes the method applicability beyond the simple analysis of structural differences in large datasets, making it a powerful tool for the inspection, interpretation and possibly generation of reactive trajectories between metastable states. Indeed, the ability to discriminate with a high level of detail different metastable states, together with the intrinsic differentiability of neural networks, make the encoded variables promising low-dimensional CVs for biased MD simulations.

The excellent results obtained for metal nanoclusters, for which the method could learn to identify a variety of structures ranging from crystalline to faulted and amorphous, demonstrates the virtue of machine learning from radial distribution functions. Building on the generality of its descriptors, this machine learning framework could be used to chart the structural landscape of diverse kinds of systems including non-metallic nanoparticles, \cite{johnston2002,anker2022extracting} colloidal assemblies \cite{Entropydrivenformationoflargeicosahedralcolloidalclusters,wang2018magic,filion2019}, advancing our capability to classify, explore, and understand transitions in these systems.

\section{Methods}\label{sec11}

The original datasets we considered included hundreds of thousands of structures for each particular cluster size and type. The structures were generated through Parallel-Tempering Molecular Dynamics PTMD simulations (see the Supporting Information). Original structures were then locally minimized to discount thermal noise. In order to avoid redundancy in the data, due to duplicates in the locally minimized structures, the initial set of structures was filtered out in order to  only select unique samples. This selection was based on both CNA classification and  potential energy. 
As a result, structures in the final dataset differed from each other by at least 0.1 meV in the potential energy or by CNA label, leading to a reduction in the number of structures to few tens of thousands for every cluster type.
The RDFs of each configurations were obtained using kernel density estimation on the interatomic distances (using the KernelDensity library from scikit-learn package \cite{scikit-learn}) with gaussian kernels and a bandwidth of 0.2 nanometers. 

The RDFs were then discretized and processed by the autoencoder as described in Fig.~\ref{fig:CNNsketch}. 
Input and output of the AE share the same sizes, equal to the total mesh points of the discretized RDFs. The convolutional part of the encoder is composed by 5 blocks made of  a convolutional layer, a rectified linear unit activation function and a batch normalization. After the convolutions the outputs are flattened and fed to a fully  connected linear layer which outputs the 3 CVs values, closing the encoder section. The decoder follows, mirroring the encoder. The 3 outputs of the encoder are fed to a another fully connected layer whose output is reshaped and fed to 5 deconvolutional blocks that replicate, mirrored, the convolutional part of the encoder. Finally, in the output layer of the decoder, data are returned to their initial size.

The output is compared to the input in the training using MSE loss. More details regarding the AE architecture parameters and the training can be found in the Supporting Information. 
After the training, the three dimensional output of the bottleneck is evaluated for all the data to obtain a 3D chart, \emph{e.g.} the one reported in Fig.~\ref{fig:chartau90}. 
After the chart of the data has been generated, the mean shift\cite{MeanShift} clustering technique is exploited to identify families of structures and evaluate the quality of the chart.  
Mean shift requires to set only one parameter, the bandwidth, dictating the resolution of the analysis.  Bandwidth selection was obtained looking for intervals of values, yielding an (almost) constant number of clusters, see Fig. S3. 

Finally, the 50 configurations closer to each centroid were analyzed visually, in order to inspect for major structural feature characterizing the different regions identified by the clustering.

\section{Acknowledgements}

This research is part of a project that has received funding from the European Research Council (ERC) under the European Union's Horizon 2020 research and innovation programme (grant agreement No. 803213).
This work has been supported by the project “Understanding and Tuning FRiction through nanOstructure Manipulation (UTFROM)” funded by MIUR Progetti di Ricerca di Rilevante Interesse Nazionale (PRIN) Bando 2017 – grant 20178PZCB5. 

The authors acknowledge PRACE for awarding us access to Marconi100 at CINECA, Italy.

\bibliographystyle{unsrt} 
\bibliography{mainbib}

\setcounter{figure}{0}
\setcounter{section}{0}

\makeatletter
\renewcommand{\thefigure}{S\@arabic\c@figure}
\makeatother

\part*{Supporting Information}

\section{Datasets}
\subsection{Filtering of the structures}
To avoid duplicates of structures that could bias the training of the network and the following clustering, we filtered the PTMD datasets, such that every structures differed from the other for at least 0.1 meV in potential energy or for CNA classification. In the following sections are listed the datasets compositions before and after the filtering, following the CNA classification of the structures.
\subsection{Au$_{90}$ dataset composition}
\begin{lstlisting}
PTMD data

Dh data = 15,520
Ih data = 789
Twin data = 83,471
Fcc data = 142,286
Mix data = 406,164
Amorphous data = 273,370

Total = 921,600
\end{lstlisting}
\begin{lstlisting}
Filtered data

Unique Dh = 1,431
Unique Ih = 766
Unique Twin = 5,946
Unique Fcc = 495
Unique Mix = 19,946
Unique Amorphous= 20,432

Total = 49,016 
\end{lstlisting}
\subsection{Au$_{147}$ dataset composition}
\begin{lstlisting}
PTMD data

Dh configurations = 278,405 
Ih configurations = 28,911 
Twin configurations = 29,553 
Fcc configurations = 19,248 
Mix configurations = 69,641 
Amorphous configurations = 188,114 

Total = 613,872
\end{lstlisting}
\begin{lstlisting}
Filtered data

Dh = 11,873 
Ih = 10,935 
Twin = 8,839 
Fcc = 2,689 
Mix = 26,615 
Amorphous = 26,099

Total = 87,050
\end{lstlisting}

\subsection{Ag$_{147}$ dataset composition}
\begin{lstlisting}
PTMD data

Dh configurations = 1,170 
Ih configurations = 117,844 
Twin configurations = 671 
Fcc configurations = 1 
Mix configurations = 5,871 
Amorphous configurations = 53,643

Total = 179,200
\end{lstlisting}
\begin{lstlisting}
Filtered data

Dh = 712 
Ih = 3,065 
Twin = 614 
Fcc = 1  
Mix = 5,249 
Amorphous = 20,764

Total = 30,405
\end{lstlisting}

\subsection{Cu$_{147}$ dataset composition}
\begin{lstlisting}
PTMD data

Dh configurations = 579 
Ih configurations = 180,260 
Twin configurations = 734 
Fcc configurations = 1 
Mix configurations = 6,422 
Amorphous configurations = 36,004

Total = 180,260
\end{lstlisting}
\begin{lstlisting}
Filtered data

Dh = 467 
Ih = 2,409 
Twin = 691 
Fcc = 1 
Mix = 5,670 
Amorphous = 18,619

Total = 27,857
\end{lstlisting}

\section{Autoencoder}
\subsection{Inputs}
\begin{figure}
        \centering
        \includegraphics[width= 1.0\textwidth]{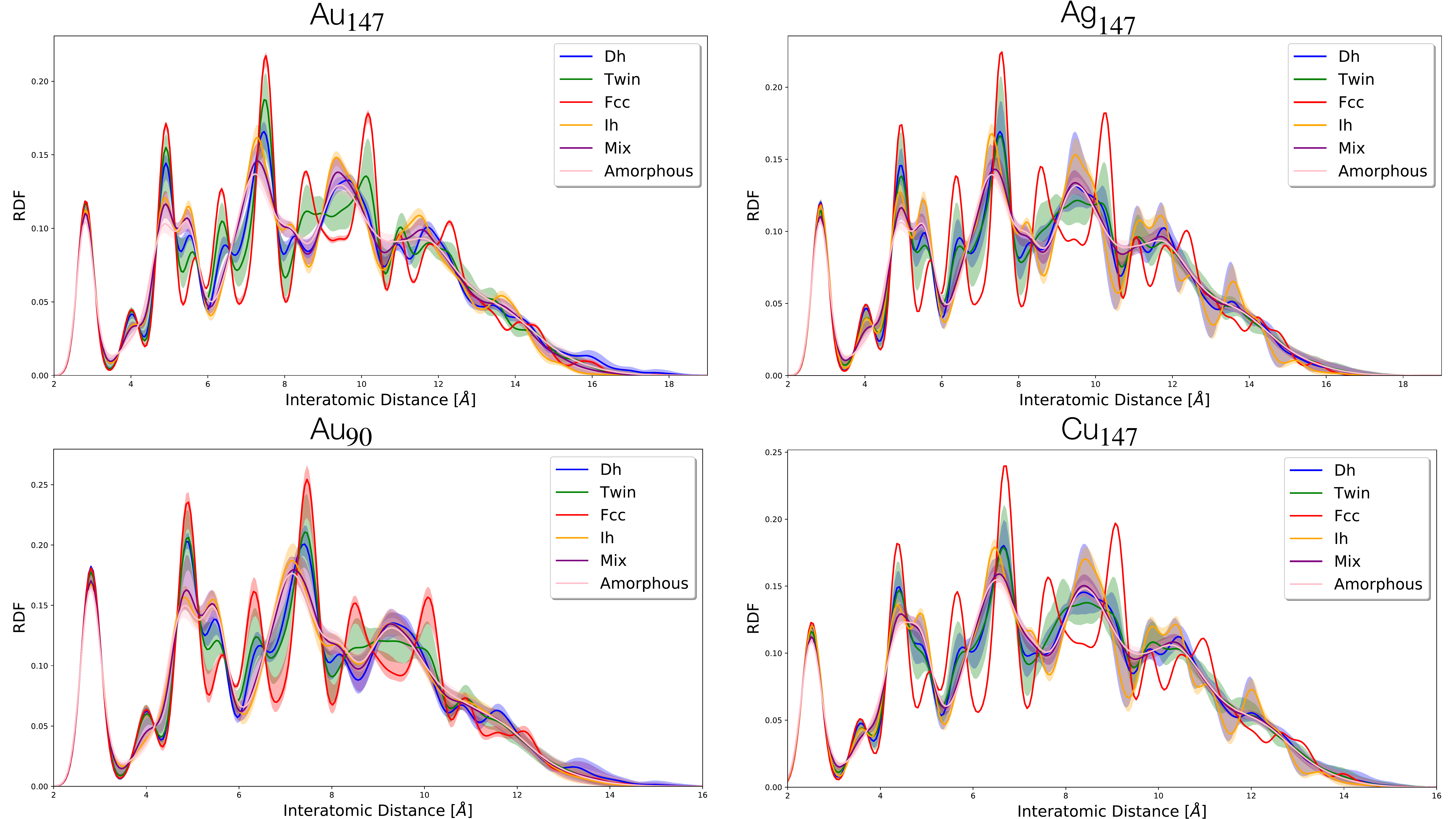}
        \caption{Plots for all the 4 metal clusters studied of the RDFs used as input to the AE, grouped following the CNA classification. Shaded regions represent the scattering of the data included between the 0.95 and 0.05 quantiles.}
        \label{fig:rdfall}
\end{figure}

The input of the AE are sequences of values representing the discretization RDFs of the structures composing the datasets. These have been obtained using kernel density estimation (using KernelDensity library from scikit-learn package \citeSM{scikit-learn}) on the interatomic distances of the single structure in order to obtain a smoothed histogram. The kde have been applied using gaussian kernels and a bandwidth of 0.2. The kde have been then fitted on the distances ranging from 0 to 25 angstrom, discretized in 500 bins, for all the metal nanoclusters studied. 
Before feeding the RDFs to the CNN autoencoder, they have been subsequently cleaned from the points where their value was identically zero for all the data in the dataset, corresponding to highest or very low distances.
The final form of the input descriptors is then a kernel density estimation of the interatomic distances discretized in 340 points, ranging from 2.004 to 18.988 angstrom for Au$_{147}$, Ag$_{147}$, while for Au$_{90}$ and Cu$_{147}$, due to the smaller sizes of the particles, only 280 points have been fed to the CNN, ranging from ranging from 2.004 to 15.982 angstrom.
In Fig. \ref{fig:rdfall} in the different panels are plotted the RDFs for all the clusters we analyzed.
All the dataset have been fed to the autoencoders splitting them in training and validation set, following the same proportion: 80\% of the data composed the training set and 20\% the validation set. The training set has been then divided in batches of 128 elements.
\subsection{Dimensionality of the latent space}
\begin{figure}
        \centering
        \includegraphics[width= 1.0\textwidth]{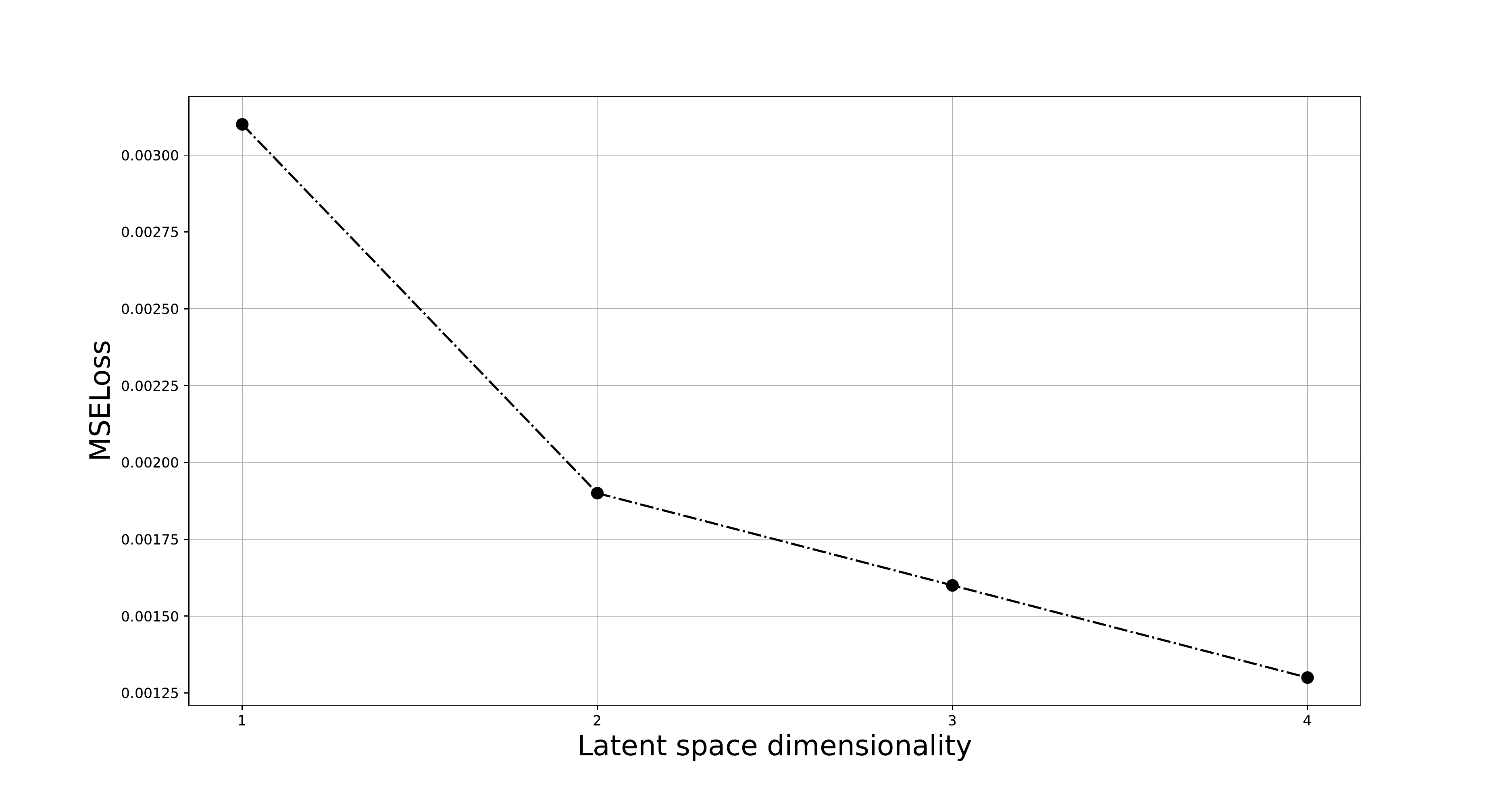}
        \caption{Plots of the MSELoss achieved after 100 epochs of training for the Au$_{147}$ dataset. All the training where conduced changing only the bottleneck size of the AE, and letting untouched all the other parameters regarding the AE structure and the training. We chose a bottleneck size, i.e the latent space dimensionality, equal to 3, in order to achieve the best performances in the training while having a visualizable and easier to analyze chart.}
        \label{fig:lossvsdim}
\end{figure}

In order to choose the dimensionality of charts, i.e. the size of the bottleneck layer of the AE, we monitored the changes in the performances of the AE from the loss point of view, changing the bottleneck size and keeping all the other parameters fixed, either in the AE structure either in the training. The best values for the loss achieved for the particular case of Au$_{147}$ after 100 epochs of training are reported in Fig. \ref{fig:lossvsdim} \\
The performances, as showed in the plot, emprove increasing the number of the latent space dimensions. In particular there is an elbow for a dimensionality equal to 2. The final choice was to pick the dimensionality that gave for better performances while still allowing for an easy visualization and analysis of the latent space. For all these reasons the final choice was a latent space with dimensionality equal to 3. The choice was supported by the good results given studying the obtained space via  a clustering technique, followed by the structural analysis.
\subsection{Structure of the autoencoder}
As described in the Methods Section of the paper, the autoencoder is composed by two main blocks, the encoder and the decoder composed by 5 convolutional layers, and a central block composed by fully connected layers. Input and output layer share the same structures and are two convolutional layers.   

The convolutional layers of the encoder share the same padding, set to 0, and same stride, set to 1. The input layer is a convolution channel with 1 input channel and 64 output channels. The number of input and output channels is the same for the next three layers, equal to 64, while the last two layers give as output a decreasing number of channels, respectively 32 and 16, while the input number of channels is set to be equal to the output of the previous layer, respectively 64 and 32. The kernels have a decreasing size, being respectively of size 20,20,20,10,8,5.  All convolutional layers are followed by a ReLU activation function and by a layer of batch normalization.
The outputs of the final 16 channels, reaching the central block, are flattened and then passed to a fully connected layer which passes them to the bottleneck, the central layer composed by three nodes, with no activation function interposed. The bottleneck is followed by a fully-connected layer of the same size of the one preceding it, then the output is reshaped and passed to the decoder, which has a completely mirrored structure respect to the encoder, with the difference that the layers apply a deconvolution instead of a convolution, in order to bring back the data to their original size.
Here is reported the summary of the AE used for Au$_{90}$ as printed by torchsummary:

\begin{lstlisting}[caption= Summary of the network printed with torchsummary]
##### [NN Architecture] #####

Autoencoder(
  (input_layer): Conv1d(1, 64, kernel_size=(20,), stride=(1,))
  (relu_1): ReLU()
  (batchnorm): BatchNorm1d(64, eps=1e-05, momentum=0.1, affine=True, track_running_stats=True)
  (encoder): Sequential(
    (0): Conv1d(64, 64, kernel_size=(20,), stride=(1,))
    (1): ReLU()
    (2): BatchNorm1d(64, eps=1e-05, momentum=0.1, affine=True, track_running_stats=True)
    (3): Conv1d(64, 64, kernel_size=(20,), stride=(1,))
    (4): ReLU()
    (5): BatchNorm1d(64, eps=1e-05, momentum=0.1, affine=True, track_running_stats=True)
    (6): Conv1d(64, 64, kernel_size=(10,), stride=(1,))
    (7): ReLU()
    (8): BatchNorm1d(64, eps=1e-05, momentum=0.1, affine=True, track_running_stats=True)
    (9): Conv1d(64, 32, kernel_size=(8,), stride=(1,))
    (10): ReLU()
    (11): BatchNorm1d(32, eps=1e-05, momentum=0.1, affine=True, track_running_stats=True)
    (12): Conv1d(32, 16, kernel_size=(5,), stride=(1,))
    (13): ReLU()
    (14): BatchNorm1d(16, eps=1e-05, momentum=0.1, affine=True, track_running_stats=True)
  )
  (maxpooling): MaxPool1d(kernel_size=2, stride=2, padding=0, dilation=1, ceil_mode=False)
  (flatten): Flatten()
  (embed_linear): Linear(in_features=3248, out_features=3, bias=True)
  (decode_linear): Linear(in_features=3, out_features=3248, bias=True)
  (decoder): Sequential(
    (0): ConvTranspose1d(16, 32, kernel_size=(5,), stride=(1,))
    (1): ReLU()
    (2): BatchNorm1d(32, eps=1e-05, momentum=0.1, affine=True, track_running_stats=True)
    (3): ConvTranspose1d(32, 64, kernel_size=(8,), stride=(1,))
    (4): ReLU()
    (5): BatchNorm1d(64, eps=1e-05, momentum=0.1, affine=True, track_running_stats=True)
    (6): ConvTranspose1d(64, 64, kernel_size=(10,), stride=(1,))
    (7): ReLU()
    (8): BatchNorm1d(64, eps=1e-05, momentum=0.1, affine=True, track_running_stats=True)
    (9): ConvTranspose1d(64, 64, kernel_size=(20,), stride=(1,))
    (10): ReLU()
    (11): BatchNorm1d(64, eps=1e-05, momentum=0.1, affine=True, track_running_stats=True)
    (12): ConvTranspose1d(64, 64, kernel_size=(20,), stride=(1,))
    (13): ReLU()
    (14): BatchNorm1d(64, eps=1e-05, momentum=0.1, affine=True, track_running_stats=True)
  )
  (output_layer): ConvTranspose1d(64, 1, kernel_size=(20,), stride=(1,))
)
----------------------------------------------------------------
        Layer (type)               Output Shape         Param #
================================================================
            Conv1d-1              [-1, 64, 261]           1,344
              ReLU-2              [-1, 64, 261]               0
       BatchNorm1d-3              [-1, 64, 261]             128
            Conv1d-4              [-1, 64, 242]          81,984
              ReLU-5              [-1, 64, 242]               0
       BatchNorm1d-6              [-1, 64, 242]             128
            Conv1d-7              [-1, 64, 223]          81,984
              ReLU-8              [-1, 64, 223]               0
       BatchNorm1d-9              [-1, 64, 223]             128
           Conv1d-10              [-1, 64, 214]          41,024
             ReLU-11              [-1, 64, 214]               0
      BatchNorm1d-12              [-1, 64, 214]             128
           Conv1d-13              [-1, 32, 207]          16,416
             ReLU-14              [-1, 32, 207]               0
      BatchNorm1d-15              [-1, 32, 207]              64
           Conv1d-16              [-1, 16, 203]           2,576
             ReLU-17              [-1, 16, 203]               0
      BatchNorm1d-18              [-1, 16, 203]              32
          Flatten-19                 [-1, 3248]               0
           Linear-20                    [-1, 3]           9,747
           Linear-21                 [-1, 3248]          12,992
  ConvTranspose1d-22              [-1, 32, 207]           2,592
             ReLU-23              [-1, 32, 207]               0
      BatchNorm1d-24              [-1, 32, 207]              64
  ConvTranspose1d-25              [-1, 64, 214]          16,448
             ReLU-26              [-1, 64, 214]               0
      BatchNorm1d-27              [-1, 64, 214]             128
  ConvTranspose1d-28              [-1, 64, 223]          41,024
             ReLU-29              [-1, 64, 223]               0
      BatchNorm1d-30              [-1, 64, 223]             128
  ConvTranspose1d-31              [-1, 64, 242]          81,984
             ReLU-32              [-1, 64, 242]               0
      BatchNorm1d-33              [-1, 64, 242]             128
  ConvTranspose1d-34              [-1, 64, 261]          81,984
             ReLU-35              [-1, 64, 261]               0
      BatchNorm1d-36              [-1, 64, 261]             128
  ConvTranspose1d-37               [-1, 1, 280]           1,281
================================================================
Total params: 474,564
Trainable params: 474,564
Non-trainable params: 0
----------------------------------------------------------------
Input size (MB): 0.00
Forward/backward pass size (MB): 3.18
Params size (MB): 1.81
Estimated Total Size (MB): 4.99
----------------------------------------------------------------
\end{lstlisting}

\subsection{Training}
The autoencoder has been trained using MSE loss function and Adam optimizer \citeSM{adam}. The starting learning rate has been set to 0.001 and then updated using a step scheduler halving its value at epoch number 30 and 90. The training was interrupted when the loss started to reach a plateau and when the obtained charts started to show no differences. All the autoencoders showed a convergence in the loss value and in the generated chart after about 100 epochs. The model showing the best loss on validation set was saved during the training.
In the table below are listed the best values for the loss for the 4 studied metal nanoclusters.
\begin{center}
\begin{tabular}{ || c | c || }
\hline
& Best loss (MSE) \\
\hline
Au$_{90}$ & 0.0029 \\
\hline
Au$_{147}$ & 0.0016 \\
\hline
Ag$_{147}$ & 0.0022 \\
\hline
Cu$_{147}$ & 0.0023  \\ 
\hline
\end{tabular}
\end{center}

\section{Clustering (Mean shift)}
\subsection{Bandwidth Selection}
\begin{figure}
        \centering
        \includegraphics[width= 1.0\textwidth]{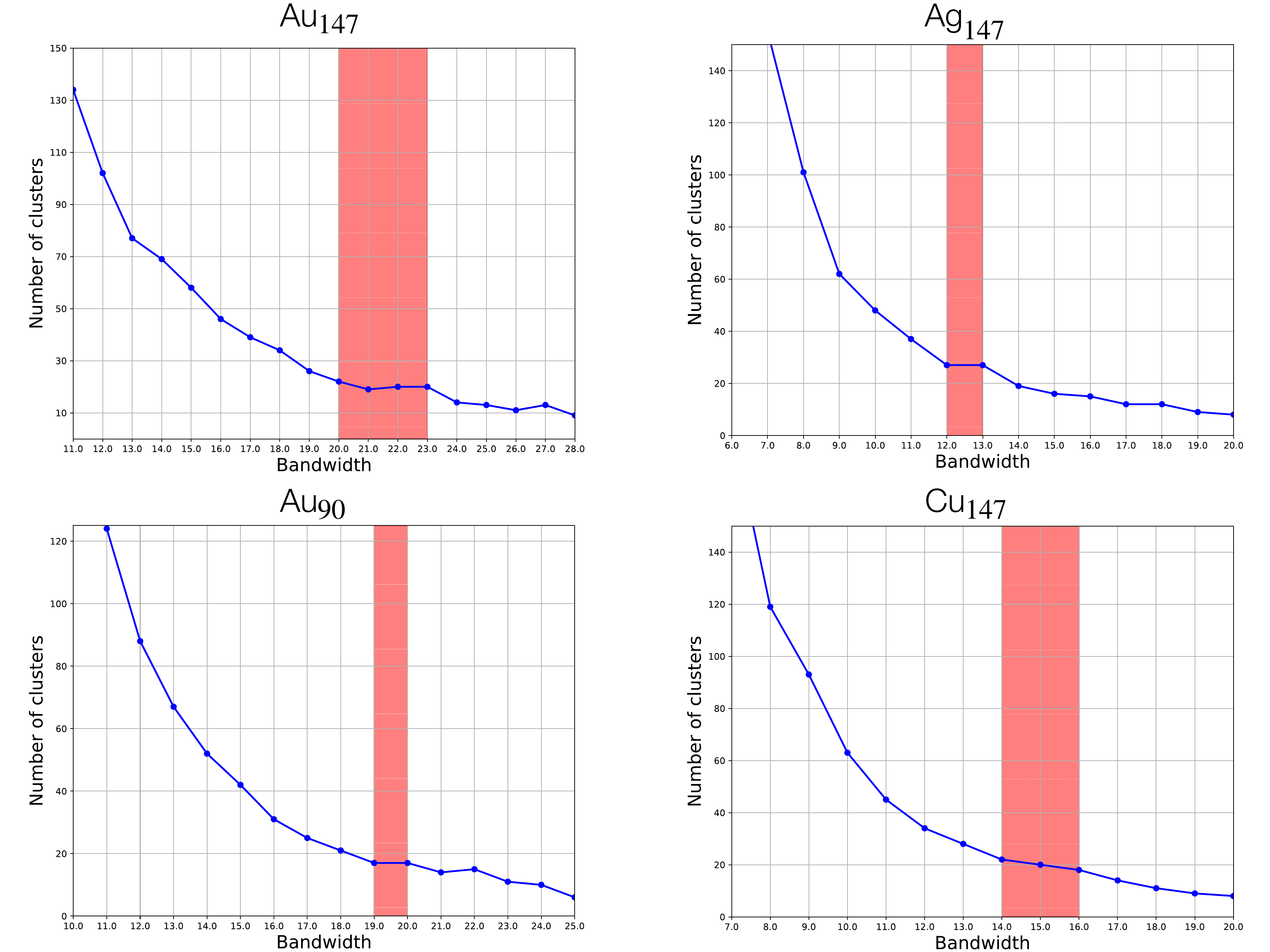}
        \caption{Plots of number of clusters as function of the MS bandwidth. The red shaded region correspond to the range of bandwidth values were the number of clusters of the decomposition shows very low dependency on the bandwidth value, indicating a good stability of the decomposition.}
        \label{fig:ms_metric_au147}
\end{figure}

The study of the reduced three-dimensional space was conduced using clustering techniques in order to locate interesting regions representing different categories of structures. Since the data and the structure of the reduced space were inherently highly inhomogeneous, we decided to exploit a technique suitable for clustering data in different sized and shaped cluster. We chose to use mean shift \citeSM{MeanShift}. This technique requires only to set a single parameter which is the bandwidth, related to the resolution of the analysis of the space.
Different techniques can be used to select the proper bandwidth \citeSM{MeanShift}. We decided to select the bandwidth among a range of bandwidth that offered a good stability of the decomposition, i. e. there was a poor dependence of the number of clusters obtained on the bandwidth value (Fig. \ref{fig:ms_metric_au147}) and to later visualize the decomposition to decide whether it was consistent with the structure of the data. 

\section{Results for other metallic nanoclusters}
In the following subsections are reported figures with the complete charting for the other clusters studied, apart from Au$_{90}$ presented in the main text.
\subsection{Au$_{147}$}
\begin{figure}
        \centering
        \includegraphics[width= 1.\textwidth]{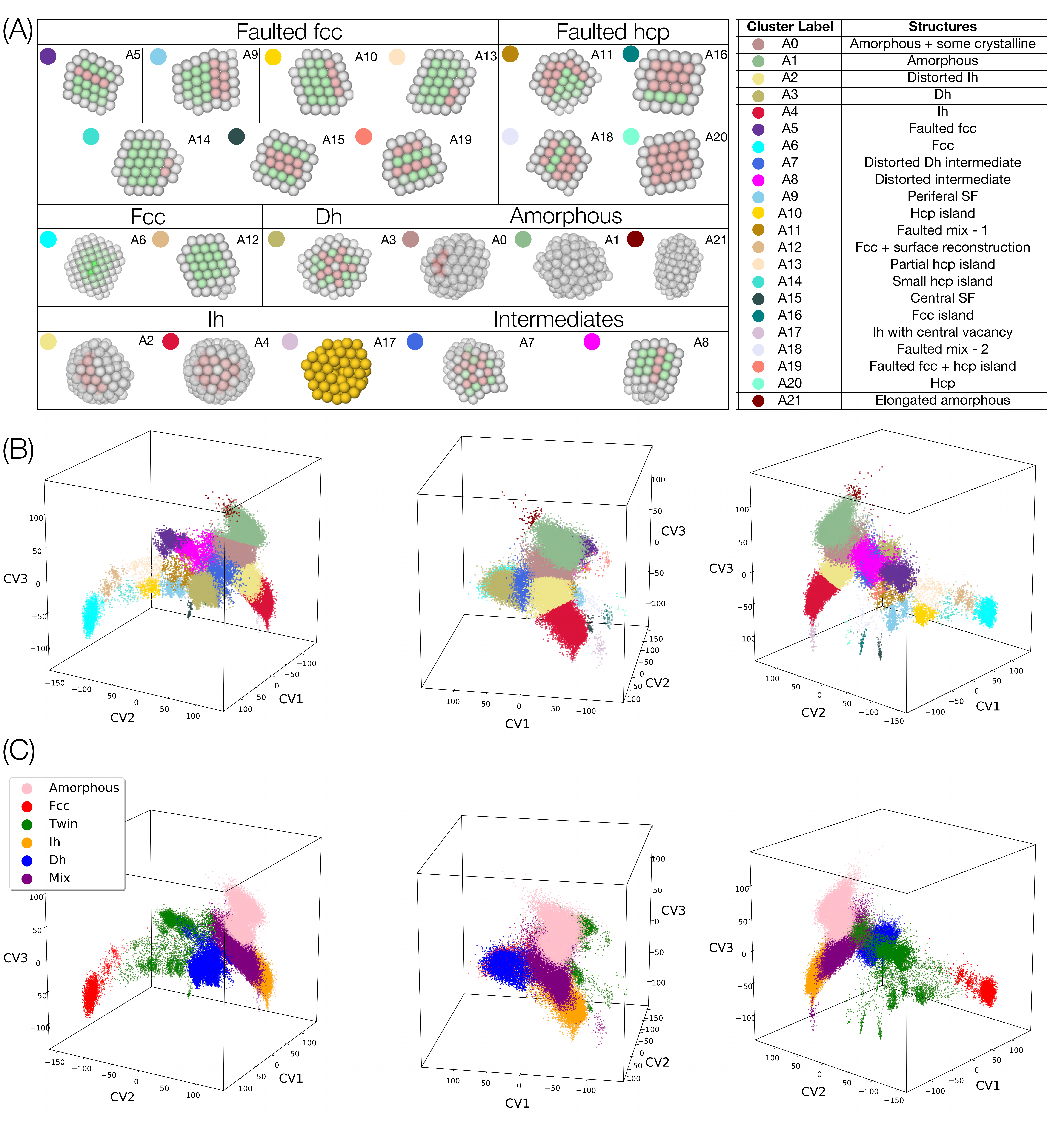}
        \caption{
        A) Au$_{147}$ figures of the 22 main structural families identified via mean shift clustering. Every figure is associated to an alpha numeric label and a color. In the table on the right are reported the descriptions of each structural family. 
        B) Different perspectives of the 3D chart of Au$_{147}$. Colors follow the mean shift clustering labels, as in panel A.
        C) Same chart of panel B, where colors now represent the CNA classification of the configurations.}
        \label{fig:chartsau147}
\end{figure}

Chart for Au$_{147}$ is reported in Fig. \ref{fig:chartsau147} with a detailed description of the families of structures identified using mean shift.
\subsection{Ag$_{147}$}
\begin{figure}
        \centering
        \includegraphics[width= 1.\textwidth]{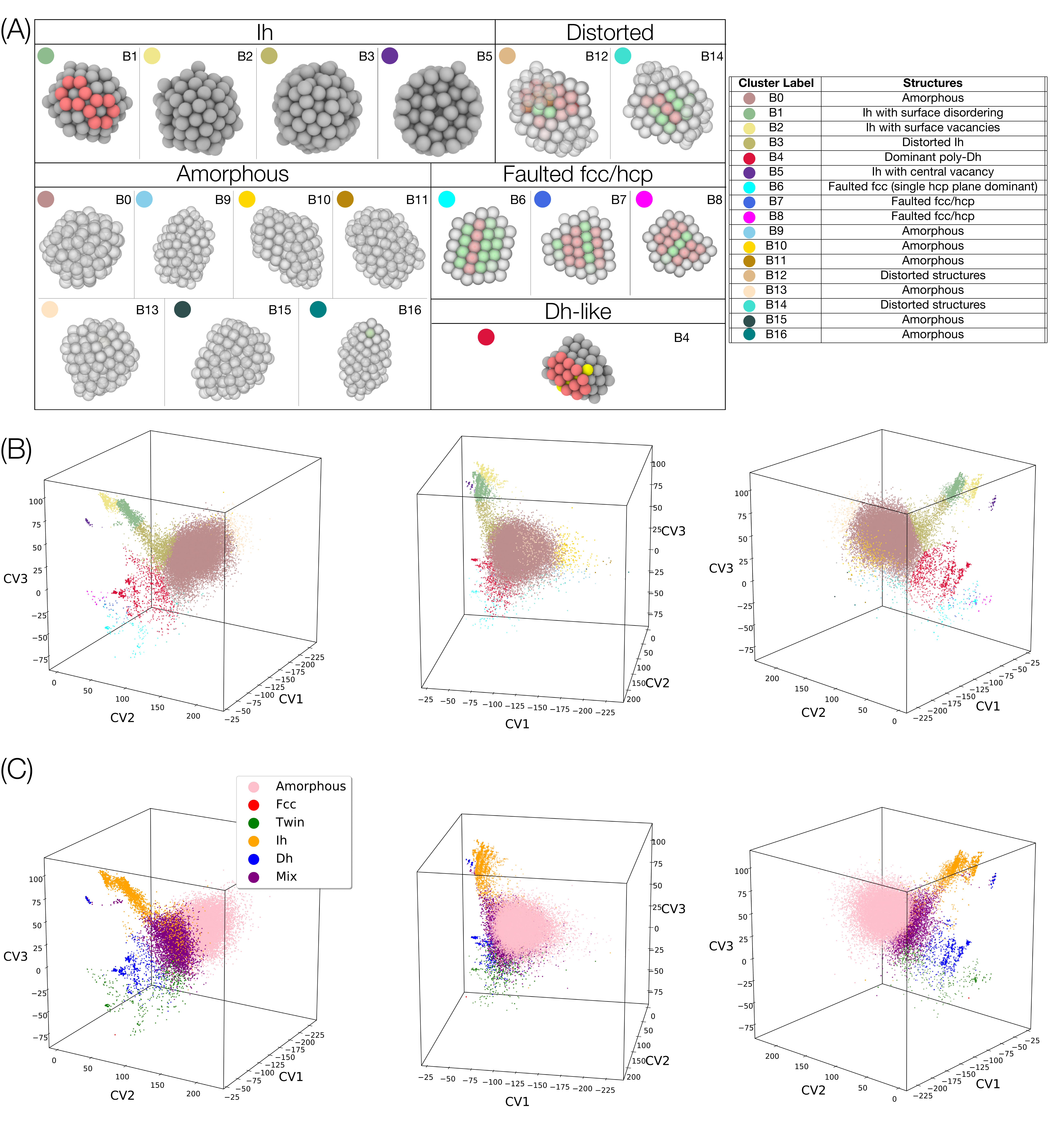}
        \caption{
        A) Ag$_{147}$ figures of the 17 main structural families identified via mean shift clustering. Every figure is associated to an alpha numeric label and a color. In the table on the right are reported the descriptions of each structural family. 
        B) Different perspectives of the 3D chart of Ag$_{147}$. Colors follow the mean shift clustering labels, as in panel A.
        C) Same chart of panel B, where colors now represent the CNA classification of the configurations.}
        \label{fig:chartsag147}
\end{figure}

Chart for Ag$_{147}$ is reported in Fig. \ref{fig:chartsag147}. Ag$_{147}$ and Cu$_{147}$ are very similar, having both a structural landscape dominated by Ih structures. Other structural motifs are few and configurations tends to be quite different between them, leading to a very scattered representation of those regions. 
\subsection{Cu$_{147}$}
\begin{figure}
        \centering
        \includegraphics[width= 1.\textwidth]{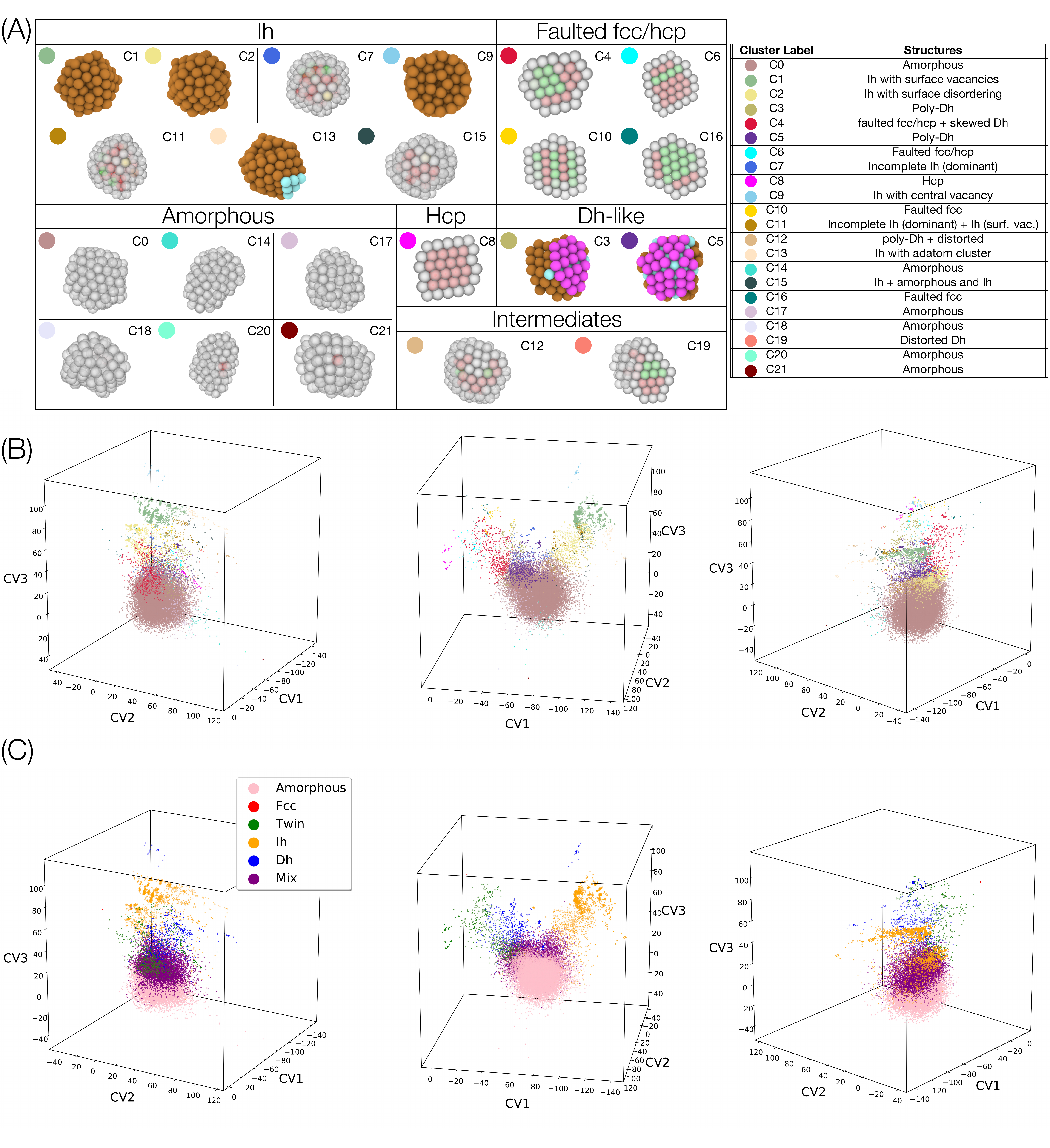}
        \caption{
        A) Cu$_{147}$ figures of the 22 main structural families identified via mean shift clustering. Every figure is associated to an alpha numeric label and a color. In the table on the right are reported the descriptions of each structural family. 
        B) Different perspectives of the 3D chart of Cu$_{147}$. Colors follow the mean shift clustering labels, as in panel A.
        C) Same chart of panel B, where colors now represent the CNA classification of the configurations.}
        \label{fig:chartscu147}
\end{figure}

Chart for Cu$_{147}$ is reported in Fig. \ref{fig:chartscu147}.

\section{Dynamical transition}
\subsection{CVs plots during main transitions}
\begin{figure}
        \centering
        \includegraphics[width= 1.\textwidth]{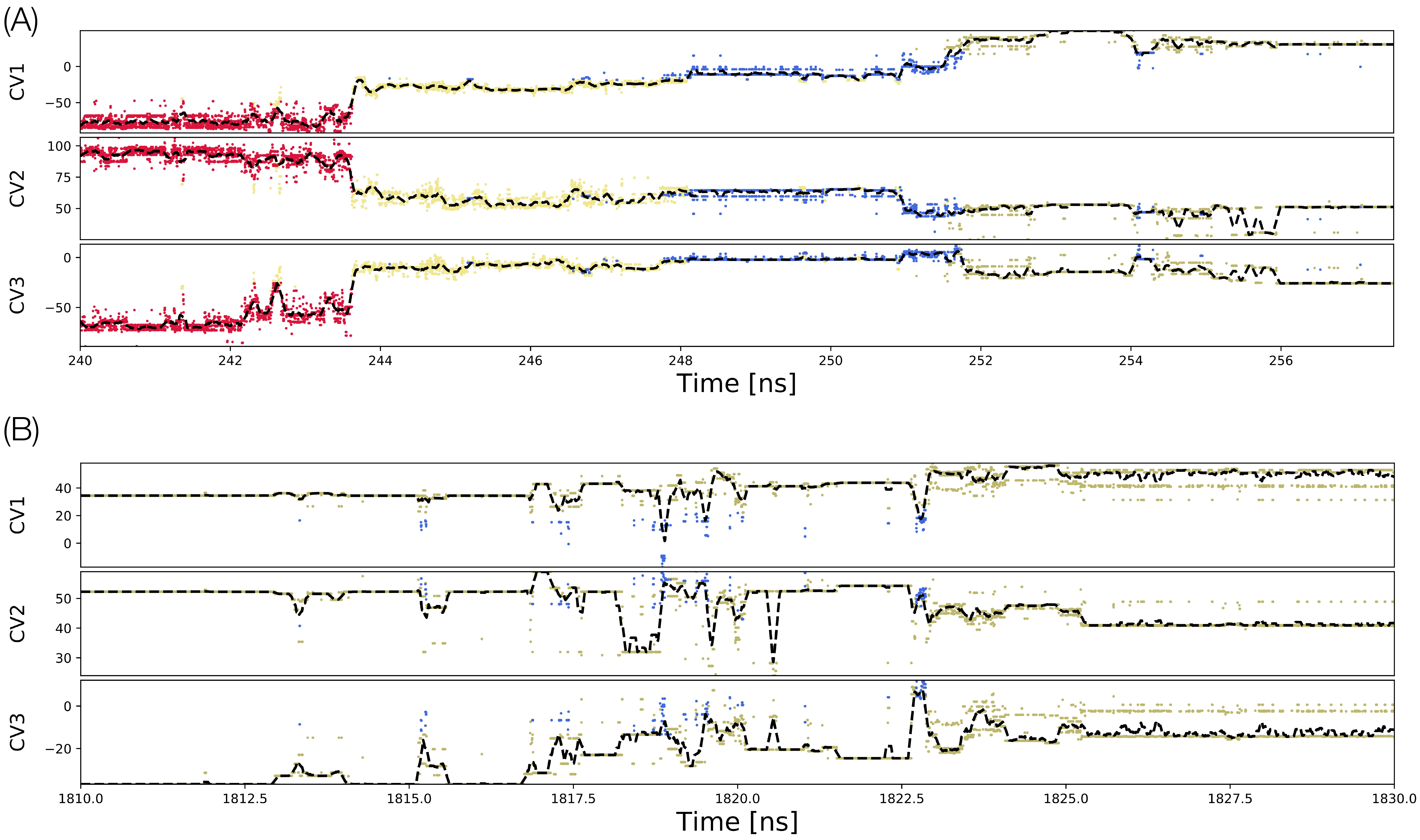}
        \caption{
        A) Plot of the 3 CVs versus time during the first main transition from Ih to Dh with an hcp island. 
        B) Plot of the 3 CVs versus time during the second main transition from a Dh with hcp island to a better Dh (with no hcp island). Colors in both panels refer to the mean shift labels of Au$_{147}$, reported in Fig. \ref{fig:chartsau147}.}
        \label{fig:cvs}
\end{figure}

In Fig. \ref{fig:cvs} are reported the plots for all the three CVs in the regions highlighted in Fig. 6 of the main text.

\FloatBarrier

\bibliographystyleSM{unsrt}
\bibliographySM{sibib}

\end{document}